# Rebuilding the Habitable Zone From the Bottom Up with Computational Zones

**Caleb Scharf**

NASA Ames Research Center, Moffett Field, California, USA.

caleb.a.scharf@nasa.gov

**Olaf Witkowski**

Cross Labs, Cross Compass Ltd., Kyoto, Japan

College of Arts and Sciences, University of Tokyo, Tokyo, Japan

olaf@crosslabs.org

**Abstract**

Computation, if treated as a set of physical processes that act on information represented by states of matter, encompasses biological systems, digital systems, and other constructs, and may be a fundamental measure of living systems. The opportunity for biological computation, represented in the propagation and selection-driven evolution of information-carrying organic molecular structures, has been partially characterized in terms of planetary habitable zones based on primary conditions such as temperature and the presence of liquid water. A generalization of this concept to *computational zones* is proposed, with constraints set by three principal characteristics: capacity (including computation rates), energy, and instantiation (or substrate, including spatial extent). Computational zones naturally combine traditional habitability factors, including those associated with biological function that incorporate the chemical milieu, constraints on nutrients and free energy, as well as element availability. Two example applications are presented by examining the fundamental thermodynamic work efficiency and Landauer limit of photon-driven biological computation on planetary surfaces and of generalized computation in stellar energy capture structures (a.k.a. Dyson structures). It is suggested that computational zones involving nested structures or substellar objects could manifest unique observational signatures as cool far-infrared emitters. While these latter scenarios are entirely hypothetical, they offer a useful, complementary, introduction to the potential universality of computational zones.

# 1. Introduction

## 1.1 Zones of habitability and computation

Despite the understood complexity of the concept of "habitability" in astrobiology and exoplanetary science (e.g., Cockell et al., 2016, Ramirez, 2018), the model of a circumstellar habitable zone (HZ, Kasting et al., 1993, and historically Newton, 1678; Maunder, 1913; Huang, 1966) is often the default approach used to evaluate the potential for life on rocky exoplanets. The HZ is motivated by the observation that most surface or near-surface dwelling organisms on the Earth seem to operate within environmental conditions roughly coincident with the natural occurrence of water in a liquid state. Liquid water's role as a key biochemical solvent, reactant, product, and catalyst is also well established. The same conditions that allow for liquid water also ensure the relative stability of biomolecules and, as we now understand, thermal (cf. Brownian) motion and ratcheting that plays a key role in driving the function of large biomolecules (Spirin, 2009), including those directly involved in the replication, translation, and transcription of information in DNA and RNA (i.e., polymerases, ribosomes). Altogether these factors appear to strongly constrain the ideal thermal and energetic environment for terrestrial type biochemistry and support the general idea of a HZ. It should be noted however that selective pressures have (from the earliest times) presumably driven terrestrial life to function this way, and alternative biochemistries in different solvent and thermal conditions are not ruled out (e.g., Bains, 2004, Schulze-Makuch and Irwin, 2018, Lingam and Loeb, 2021).

Moist greenhouse effects and water loss at the hot "inner" HZ as well as high albedo $CO_2$ condensation effects at the "cold" outer HZ are typically incorporated into estimating HZs (Kasting, 1993, Kopparapu et al., 2013). However, the calculation of the orbital range of a HZ is in almost all instances approximate owing to the dependency on many factors that are typically unknown or poorly constrained, such as: atmospheric composition, spin-orbital configuration, land-ocean distributions, geophysical conditions, and gravitational tidal heating. Key efforts to address some of these factors have studied the addition of $H_2$ and $CH_4$ to planetary atmospheres already containing greenhouse gasses like $CO_2$ and $H_2O$ (e.g., Pierrehumbert and Gaidos, 2011; Ramirez and Kaltenegger, 2017, 2018) and tidal effects (e.g., Barnes et al., 2009). The fidelity of models is also a factor, and, at best, modeling involves the application of full 3D general circulation models (GCMs) of rocky planet climate states that still necessarily simplify physics and chemistry in aid of numerical efficiency (e.g., Way et al. 2017). Because of stellar evolution along the hydrogen-fusing main sequence, the HZ is also entirely time dependent (e.g., Kasting et al., 1993, Ramirez and Kaltenegger, 2014), with factors such as geophysical activity and the carbon-silicate cycle, (Kasting et al., 1993) contributing to that dependency.

Application of the HZ is also quite often disconnected from the other known requirements of, or opportunities for, living systems. Significant work has been undertaken to address this, with more nuanced, generalized evaluations of habitability (to environments such as those of the interior oceans of icy worlds) by focusing instead on the detailed conditions of free energy availability to organisms and the nature of metabolic pathways (e.g. Nealson, 1997; Hoehler, 2007; Hoehler et al., 2007, Cockell et al. 2016) as well as the availability of rarer elements such as phosphorus or molybdenum that are essential for modern terrestrial biomolecule structural elements and cellular energy transport (phosphorus in DNA and Adenosine Triphosphate - ATP), as well as enzymatic processes (molybdenum in nitrogenase enzymes). Some of these efforts have progressed to enable sophisticated models of organism populations in 'extreme' conditions that show promise for establishing more biochemical metrics for habitability (Higgins and Cockell, 2020). Other work has been undertaken to build a definition of a HZ for complex life (Schwieterman et al., 2019, Ramirez, 2020).

At the same time, new ideas and speculative principles on the universal possibilities for, and observable consequences of, life (including life that might utilize different substrates) have continued to develop. For example, see Kempes and Krakauer (2021) for a multi-path theory of origins of life, or the assembly theory approach (e.g. Marshall et al., 2021; Sharma et al., 2022), or various uses (e.g., Zenil et al., 2023) of algorithmic information theory (Chaitin, 2004). These include the possibility that new, naturally selected, evolving systems might be constructed or emerge inadvertently from, or in tandem with, biological life (e.g., Scharf, 2021; Frank et al., 2022). Driving many of these concepts is the idea that living systems are innately informational in nature; that self-propagating information (in DNA and RNA sequences, together with many peripheral compounds, as well as transient electrochemical signals) is the only truly persistent phenomenon, while the substrate structured to carry that information is as much an evolving consequence as a cause (e.g., Walker and Davies, 2013). It has been suggested that conceptually (and perhaps more literally) terrestrial organisms were the first "data scientists", owing to the striking similarity between life's assimilation and utilization of data and the procedures and principles of modern informatics (Wong and Prabhu, 2023).

Self-propagation through processes such as replication, self-repair, or predictions that enhance the possibility of future continuance, indicate a universal requirement for information to exert itself over matter, and to drive physical processes that effectively utilize self-information and relevant mutual information about the environment (e.g., Krakauer, 2011). Effective utilization of information may involve minimization of discrepancies between an internal predictive (e.g., Bayesian) model and the surrounding world. Here it is suggested (similarly to others, e.g., Kempes et al., 2017; Kolchinsky and Wolpert, 2018) that this feature is best labeled "computation". Although this term is more commonly used in describing a mathematical procedure or a mechanical/digital process (see below), it nonetheless seems the most appropriate label here. This conceptual framing is further illustrated by modeling of the emergence and evolution of living systems *in silico* (Bedau and Cleland, 2018), including artificial chemistries (Fontana and Buss, 1994; Banzhaf and Yamamoto, 2015), self-replicating programs competing for CPU time (Ray, 1993; Ofria and Wilke, 2004), or simulated creatures evolving complex behaviors (Sims, 1994). These models highlight the fundamentally computational character of living patterns, tied to computational measures (Bedau and Packard, 1991; Adami and Cerf, 2000).

The informational perspective, albeit an extremely powerful way of capturing causal relations in living and complex phenomena, does currently come with important caveats. There seem to be fundamental limits to measures of information in living systems, starting from Shannon information or mutual information, all the way across a growing collection of mathematical measures, up to recent theories of integration and causality. Such limits are attached to a dependency upon a fixed observer, determining the variables or degrees of freedom at play (Adami, 2016). This, in turn, points back to basic relations between information that allows accurate predictions for a given system, and thermodynamic entropy. While entropy represents unpredictability in a system, information is the reduction of uncertainty in the system (Shannon, 1948). Understanding information in systems thus not only involves quantifying the data or the mutual dependencies among variables but also entails a deeper grasp of how these systems organize and reduce complexity, thereby lowering their overall entropy. Jaynes (1965) and Adami (2016) have made clear arguments about the observer dependency for entropy and information, respectively.

Attempts have been made to incorporate the property of "meaning" in information. For example, Kolchinsky & Wolpert (2018) proposed that living systems encompass both "stored semantic information" (the mutual information between a system and its environment with scrambled probabilities in their initial condition), and "observed semantic information", the scrambled transfer entropy measuring how observations of an environment help predict the future of a system. Another idea emphasizes measures, such as functional information (Szostak, 2003; Hazen et al., 2007), that quantify the complexity of systems in the context of specific functions of the system, in contrast to other formalisms based on informational contents. Parallel efforts examine "relevant information", such as in Donaldson-Matasci et al. (2010), who studied the fitness value of information in terms of the maintenance of a system; tightly connected to the survival of living systems in uncertain environments. This captures how the fitness value of a developmental cue equals the reduction in uncertainty about the environment as described by Shannon information.

These approaches lead to informational measures of the self-maintaining properties of living systems in an environment. Information closure (Bertschinger et al., 2006) allows a distinction to be made between life and non-life, by an observer-independent characterization of entities that display autopoiesis (Maturana and Varela, 1991) (the capacity to produce or maintain oneself by building one's own parts). This bears some relation to approaches such as "empowerment", or the information flow between actions and sensory inputs (Klyubin et al., 2005), autonomy (Bertschinger et al., 2008), and individuality (Krakauer et al., 2020). It should be noted that these concepts connect fundamentally to paradigms of autonomous machines and replicators, as studied by John von Neumann (1966), and connect to strategies for – and thus technosignatures of – spacecraft for space mining and exploration (Freitas, 1980; Borgue and Hein, 2021; Matloff, 2022). Although the information flows on Earth have been dominated by the biosphere, they may be outgrown by the technosphere in about 90 years due to its rapid exponential growth (Lingam et al., 2023). This also naturally leads us to computational approaches for identifying life.

A key consequence of framing the functionality of life as computation is the availability of powerful tools from information theory that provide a fundamental connection between the state changes and manipulations of matter inherent to the instantiation of information and the thermodynamics of those changes (see below). This has led to proposals for understanding the nature of biological information (Adami, 2012; Koonin, 2016) and, as mentioned above, the quantification of *semantic* information (as opposed to syntactic or Shannon information) as (mutual) information that a living system contains about the environment that directly influences that system's persistence in the world (Kolchinksy and Wolpert, 2018). Other studies have examined (again through information theoretics, such as rate-distortion theory, Marzen and DeDeo, 2017), the compression of that semantic, sensory information in terms of the cost to a living system of acquiring and maintaining information about its environment versus the benefits as a function of the degree of information fidelity or resolution. Additionally, algorithmic information theory (Chaitin, 2004) characterizes strings of information, built upon concepts like Kolmogorov complexity – the length of the shortest possible algorithmic description of a data string – which illustrates how such tools can help distinguish between patterns arising from natural processes and those that may be indicative of technological constructs. Finally, there may also be particularities to information dynamics in collectives (cf. Kim et al. (2021) for a review). All of the above problems are fundamentally computational.

The physical basis of this kind of computation lies in the changes in states (of bits, molecules, atoms etc.) associated with the emergence, use, and propagation of semantic information. Those state changes might encompass highly localized microscopic properties (e.g., individual covalent bond making or breaking, quantum spin, electrical charge etc.), as well as larger scale molecular reconfigurations or structural changes (polymerization, folding), or quantized molecular rotational or vibrational modes, translational movement, and macroscopically manifested phenomena, such as material phases, bulk property changes, relocation etc. Therefore, as applied to living systems, the label of computation is quite agnostic, it does not assume anything about how life is built, or very many details about its function. It also applies equally to life's extended physical phenotypes, which arguably include technology (Scharf, 2021).

It is proposed here that to link the many qualities of living systems, and to advance the quest to understand habitability and the probability of life in an environment, that *computational zones* (CZs, environments in which computation takes place) are a natural generalization of the idea of habitable zones. CZs provide a framework to properly combine traditional approaches to habitability: including factors such as the liquid water HZ, free energy availability, elemental and chemical availability, historical contingency, and the preexistence of living systems. Furthermore, while there is still debate as to whether classically conceived habitability is a 'binary' or 'yes, no' environmental division, or a continuum defined by binary questions (Cockell et al. 2019), or a fully continuous property (Heller 2020), computational zones may be almost indefinitely extensible, but will be modulated by energy availability and energy efficiency, along with total computational capacity as a property of the conditions of matter.

**1.2 Factors limiting computation.**

In general, computation may be defined as a universal phenomenon involving the systematic manipulation of information, following a well-defined set of rules (Wolfram 2002). Information itself can be considered as a physical entity (cf. Landauer (1991), Lloyd (2001), and Piccinini (2015)). Despite the seeming diversity of computation (e.g., across biology and technology), the past century of research on this topic has shown that all forms of computation, including Turing machines (Turing, 1937), lambda calculus (Church, 1936), and register machines (Shepherdson and Sturgis, 1963), do in fact present an equivalent set of properties (Wolfram, 2002).

However, there are still fundamental constraints that may limit computational processes from achieving certain levels of organization and 'intelligence' or reaching open-ended capabilities for continually discovering novel solutions or strategies. These constraints are imposed on computation by the laws of physics, limiting capacity, efficiency, and speed, and play a crucial role for understanding the possibilities and limitations of the use of computation by living and intelligent systems in the universe. In turn, this knowledge effectively informs the sensitivity and limits of our search for life.

In more detail: the constraints to computation range from physical memory encoding and transfer to 'hardware' cost, to purely 'software' or mathematical constraints. The first category involves spatio-temporal constraints, which are imposed by the presence of noise affecting the spatio-temporal coherence of a system, in terms of either positive (e.g., Wiesenfeld and Moss (1995)), or negative effects, such as the channel capacity of transmission without error (Shannon, 1948). Noise limits the number of factors that can be used in computation, the amount of time that can be spent before the computation will be forced to decay, and the complexity of interactions that may be chained to one another, as well as the actual storage space available.

Storage in this context is the ability of the system to implement processes of memory that can preserve correlated states in space and time (as deduced from neuroscience, e.g., Buonomano and Maass (2009)). Finally, this first category includes the basic limit of the speed of light that imposes a maximum bound on information transfers. As an extreme example, compared to a human brain, where all nodes are relatively close yet extremely limited in energy budget, is a Dyson Brain (a massive computational system built around a star), that could face communication delays and energy inefficiency due to the speed of light and energy conversion challenges (Cirkovic, 2018).

A second category of constraints arises from the properties of the physical substrate, a.k.a. the 'hardware' on which computation takes place (Papadimitriou, 1994). Even in cases where the spatio-temporal coherence is robust, the physics of the computational substrate can impose limits on what informational interactions are possible, with implications for the computability of certain functions (Piccinini, 2015; Biehl and Witkowski, 2021). For example, given the Planck and Boltzmann constants and the speed of light, Bremermann's Limit can be derived on the maximum rate of computation in an isolated physical system (Bremermann, 1962). Similarly, based on the Heisenberg uncertainty principle, the Margolus-Levitin theorem states the speed limits of information processing in quantum systems, including communication delays that limit the maximum computational speed per unit energy (Margolus and Levitin, 1998). The Bekenstein Bound provides a theoretical limit on the amount of memory that can exist in a given region of space, given a certain amount of energy (Bekenstein, 1973, 1981a, and 1981b). Energy supply itself constitutes a limit, as defined by the von Neumann-Landauer Limit for information erasure or for performing any irreversible operation on a physical system (von Neumann, 1956; Landauer, 1961; see below). It is therefore essential to consider substrate-related constraints that include factors such as the underlying physical properties of the substrate and its architecture (Marković et al., 2020, Schuman et al., 2022), its complexity (Czarnecki et al., 2020), its connectivity (Barrat and Weigt, 2000; Dale et al., 2021), and subsequent energy costs (Lloyd, 2000). All of these restrict the range of computation available to agents evolved and imposed on any substrate.

A third type of constraint arises from the 'software', or mathematics of computation, and is not strictly physical, but relates to the nature of mathematical knowledge itself (Gödel, 1931; Chaitin, 1966 and 1975; Turing, 1936). A common example is computational complexity, in space and time. That is, the inherent amount of time, memory, or any other resource needed to run some computation, as a function of the input size. This measure is itself often defined via algorithmic complexity (Kolmogorov, 1965), that refers to the length of the shortest possible description of an object that completely captures its information content and may be constrained by the underlying hardware and substrate.

Biological computation may have inherent 'software' constraints on what can catalyze or inhibit the evolution of powerful cognitive architectures, with various embodiments, in various environments (Krakauer, 2011; Kempes and West, 2019; Kempes and Krakauer, 2021). These constraints can restrict the type of informational mappings possible in each time frame (Adams et al., 2017; Walker and Davies, 2020).

In summary: the three main types of computational constraints include spatio-temporal constraints, hardware/substrate limitations, and mathematical or software constraints. Spatio-temporal constraints involve the impact of noise on computational systems, limiting factors such as storage space, and the duration and complexity of interactions. Hardware constraints are tied to the physical properties of the substrate, its architecture, complexity, connectivity, and energy costs. Finally, software or mathematical constraints encompass computational complexity, algorithmic complexity, and the constraints inherent to biological computation.

The constraints defined above are thought to be universally true and applicable across all forms of computation, from biological to technological (in the broadest sense, that might even include extremely hypothetical possibilities such as the use of cold degenerate stars or black holes, e.g., Lloyd, 2002; Susskind, 2015). These constraints must therefore also point to universal signatures and limits for the detection and characterization of living systems.

Understanding the complex interplay between all the above categories of constraints on computation in the universe is critical, but beyond the present scope. In the following sections an initial, simplifying choice is made to focus on three primary factors limiting the existence, type, and quantity of computation available or occurring in any given environment:

*1) Capacity:*
Divided into two sub-factors: 1) *The number of available states* for carrying information. For example, this could simply mean either the theoretical or practical maximum of the number of digital logic gates that can be constructed from available matter on a planet, or around a star. It could also mean the absolute number of particle states (energies and quantum mechanical properties like spin) in a volume of the universe, or, more parochially, the number of nucleotides in a biosphere's DNA or RNA and the number of components of biological function such as amino acids (see below). 2) *The maximum rate* at which those states can be manipulated in computation. In many practical instances this will have a dependency on available energy (see below), however, there will also be absolute rate limits that depend on physical properties. For example: electron transport speed and frequency limits, or molecular bond formation and breaking timescales.

*2) Energy:*

Energy available to do work that supports computation. For example: flipping bits between 1 and 0, transferring signals, and translation of stored information. Landauer (1961) introduced a general limit (or Ultimate Shannon Limit, or von Neumann-Landauer Limit) on the minimum amount of energy needed (and generated) by irreversible processes of bit erasure or the merger of computational paths: $E_{Landauer} = k_B T_0 ln\, 2$ (where $k_B$ is the Boltzmann constant, see also below). At room temperatures of $T_0 = 20°C$, $E_{Landauer}$ is about 0.02 eV ($3.2 \times 10^{-21}$J), and therefore the Landauer limit is millions of times lower than what is achieved by an average modern digital computing system. *Reversible* computation (in which no information is lost, cf. Bennett (1988)) such as momentum computing (Ray et al., 2021) would have even lower practical limits, although that may come at the cost of speed (which may or may not matter in a cosmic context), and it is not clear that momentum computing would arise as a first step for living systems in nature. The rate of computation (and hence capacity, see above) will be tied to the rate of available energy flow, but that is a strongly context dependent constraint (i.e., whether fast or slow computation is important).

*3) Instantiation/substrate:*

The ways in which computation may be carried out. In each environment there may be a range of possible mechanisms for computation to happen, and the environmental conditions will place constraints on these. Terrestrial biological computations within the framework of carbon-based molecular structures are just one example. For comparison, silicon-based digital computation presumably cannot take place at temperatures where silicon melts (approximately 1410°C for pure silicon). However, a mechanical computer built out of tungsten could operate to 3422°C. There are even more unconventional possibilities, for instance, convective cells in fluids could, in principle, be engineered to carry out computations using Boolean logic (Bartlett and Yung, 2019). Since convection is seen in environments from cold superfluids to stellar photospheres, the possibilities are diverse. Quantum computing adds to this diversity, since it can utilize a variety of qubit instantiations, from atoms to macroscopic quantum objects.

## 2. Discussion

### 2.1 Energy limits and evaluating computation.

The energy available for computation and the energetic/entropic cost of computation have fundamental constraints. For biological processes that are characterized as directly informational in nature there should be a baseline relationship to the Landauer limit on energy required/dissipated for irreversible bit operations (see above): $E_{Landauer} = k_B T_0 \ln 2$, where the system temperature $T_0$ can be equated with the environmental temperature and the natural logarithm of 2 captures the change in a binary operation. This expression arises from a consideration of the classical Boltzmann entropy ($S = k_b \ln W$) and the change in microstates ($W$) for this bit operation. $E_{Landauer}$ is considered to be the absolute limit on energy required and is thought to apply to any instantiation of information (including pure quantum states, Yan et al., 2018).

Studies on the relationship between information and computation, and changes in entropy and/or energy in biological systems have applied Landauer's principle to molecular processes (e.g., Smith, 2008; Matta and Massa, 2017). Kempes et al. (2017) examined the minimum theoretical energy requirement for an amino-acid operation, i.e., the attachment of an amino group to a polymer (protein chain) during RNA translation in a cell, since this is an explicitly information-bearing process. Estimating the Shannon entropy change in drawing amino acids (assuming a *uniform* mix of 20 unique types) from solution to form proteins with average length of 325 amino acids these authors arrive at a theoretical minimum energy for synthesizing a typical protein of $1.24 \times 10^{-20}$J per amino acid. This does not include changes in chemical binding energies, or other energetics of the molecules involved. Comparing this to the empirical determination that 4 Adenosine

Triphosphate molecules (ATP) are required (yielding an estimate of $3.17 \times 10^{-19}$ J per amino acid) it appears that the actual energy cost is $\sim 10 E_{Landauer}$ to within an order of magnitude (with a dependency on the environmental temperature).

Therefore, with that work as a guide, and assuming that the theoretical Landauer limit is an absolute constraint, the minimum energy requirement (or burden) of a computational biomolecular process that we label as $x$ can be approximated. That requirement should be $\alpha_x E_{Landauer}$ where $\alpha_x$ is a process-dependent factor. This suggests a simple, but absolute, upper limit to the number of biological operations per second ($BOPS_x$) for process $x$:

$$BOPS_x \leq \frac{1}{\alpha_x E_{Landauer}} \varepsilon_x P \quad ,$$

(1)

Where $P$ is the total power flow in the system (at the top level) and $\varepsilon_x$ is an efficiency factor that incorporates both the thermodynamic limits on power that is available to do useful work (i.e. free energy per unit time, see below) as well as any process-specific efficiency limits and partitioning of $P$ (i.e. only a fraction of power in the environment will be available to a given process). Critically, this formulation, through $E_{Landauer}$ explicitly incorporates the environmental temperature $T_0$. It should be noted (and will be explored further in future work) that $\varepsilon_x$ can be temperature dependent (see below) and that the Arrhenius-Boltzmann kinetics of chemical reaction rates indicate a rate dependency proportional to $e^{-E_a/(k_B T)}$ (where $E_a$ is the activation energy and $k_B$ is the Boltzmann constant) that could apply to the $BOPS_x$ rate.

The absolute thermodynamic efficiency and process-specific efficiency and partitioning are likely to differ. For example, while photon transduction in most photosynthesis is extremely efficient (>90%) the overall photosynthetic outcome in terrestrial plants - the conversion of water and carbon dioxide to glucose and oxygen (not an explicitly informational process, although it ultimately powers informational processes) - is known to be efficient at a level of 3-6% of the *total* power flow from stellar radiation (e.g., Zhu et al. 2008) when the wavelength range of photons capable of activating the process and the actual absorption fraction of photons for photosynthesis in plants is accounted for (plants also discard many photons and utilize photon energy in other processes such as transpiration). This is significantly lower than the thermodynamic limit on efficiency that can be estimated by approximating a stellar spectrum as an ideal blackbody of temperature $T$ and computing the energy available for useful work (also known as the exergy) given the environmental temperature $T_0$ (Scharf, 2018). Since exergy is the maximum amount of work obtainable from a system in reference to its environment, for blackbody radiation the integrated exergy $W_{T,T_0}$ for an object bathed in radiation of temperature $T$ is (e.g., Candau, 2003):

$$W_{T,T_0} = \sigma\left(T^4 - \frac{4}{3}T_0 T^3 + \frac{1}{3}T_0^4\right)$$

(2)

Therefore, the efficiency is the ratio $W_{T,T_0}/\sigma T^4$:

$$\varepsilon = 1 - \frac{4T_0}{3T} + \frac{1}{3}\left(\frac{T_0}{T}\right)^4$$

(3)

Which is typically between approximately 90% and 95% for environmental temperatures between 375K and 275K respectively and a stellar temperature of ~5800 K (Scharf, 2018). In section 2.4 below the use of stellar energy to drive computation is examined further in two different scenarios. In this present work an isotropic radiation 'bath' is assumed for simplicity. Modifications to the efficiency exist for radiation received from a finite source (e.g., a stellar object in the sky that introduces a geometric (solid angle) concentration factor $\sim (R_*/d)^2$ to the third term on the r.h.s. of Equation 3 - see for example Badescu 2023 and Wright 2023). We note however that for much phototrophic life on Earth (e.g., vascular plants) photons are sifted through multiple internal reflections before interacting with photosynthetic reaction centers, modifying the formal efficiency similarly to solar concentrators (c.f. Badescu 2023). The assumption of isotropy may therefore be a reasonable approximation for this type of phototrophic life, and in this present work we use this simplification. For the technological use of stellar radiation, where photons may be captured more directly (e.g., on one side of the surface of a solar array), the geometric concentration is likely important and isotropy cannot be assumed (e.g., Wright 2023).

Although the total efficiency factor $\varepsilon_x$ for a process like an amino-acid operation (as above) is still uncertain owing to the 'upstream' processes that are needed to go from power flow in the environment to the core operations, Equation 1 nonetheless represents an absolute upper limit to the BOPS that can be supported for a given power flow and the theoretical minimum energy requirements for bit-like operations in a specific environmental condition (i.e. via the temperature dependence of $E_{Landauer}$). Therefore, any living system, whether a single organism or a complete biosphere, can be characterized as a finite sum across the rates of all computing processes, i.e.:

$$BOPS \leq BOPS_1 + BOPS_2 + BOPS_3 + \cdots BOPS_N$$

(4)

Where the inequality holds if the general thermodynamic limits alone are applied. Here, life is being described solely via the granular biomolecular processes that involve the transfer, encoding, or erasure of information.

The identification of these informational processes may be made in a variety of ways. Specific biochemical mechanisms, such as the example of translation used above, provide one natural category. Nucleotide transcription and reverse transcription, as well as the action of regulatory processes on genes, or the disassembly of mRNA within a cell might also all form specific categories. But there are other possibilities that would clump mechanisms together under a single purpose. For instance, the processing of sensory information is an interesting example of a need for living systems that is presumed to be driven by some fundamental informational principles (e.g., lossy data compression in biology, Marzen and Dedeo, 2017). Rapid increases in environmental complexity may also drive greater 'computation' in living systems to support survival (Krakauer, 2011).

But the generation of something like a molecule of ATP is arguably *not* informational, even though its existence is a consequence of a molecular architecture stemming from the presence of information-bearing structures, and it will be involved in supporting computation. It may make the most sense to consider a hierarchy of nested biological computation (e.g,. Kempes et al., 2017), although in this present paper the focus is on the lowest level of processes.

In other words, the BOPS as described here are assumed to represent the core processes associated with replicating, evolvable systems, and life – the core complexities. This is complementary to ideas such as Assembly Theory that seeks to more rigorously, and universally, quantify the transitions between abiotic and biotic systems (Sharma, 2022). It is also complementary, albeit finer grained, than studies that use estimates of cellular activity and energy use (e.g., Falkowski et al., 2008; Bradley et al., 2020, see below) to gauge the requirements and behavior of specific components of Earth's biosphere.

This breakdown into processes, as in Equation 4, provides an alternative, nuanced way to construct a picture of environments being evaluated for the potential to harbor life. What processes, for instance, dominate the total BOPS on a terrestrial-type planet, and will these always be the same? Some computational processes may map directly to the kind of external biosignatures considered traditionally, such as atmospheric gas compositions (where species such as oxygen and methane are part of a bio-geo-aero-chemical cycle), or surface spectral features like those arising from phototrophic life. It could also be instructive to further examine what the partitioning of activity and energy supply/demand is between metabolic processes that are not computational *per se* but support essential cellular homeostasis and energy availability, and centrally important computational processes such as transcription and translation. This computational and/or energy hierarchy may change with different organism complexity or different biosphere complexity (e.g., microbes versus multicellular systems and low-energy niche biospheres versus high-energy biospheres, e.g., Hoehler, 2007; Higgins and Cockell 2020).

**2.2 Fraction of instances: habitability and occupancy**

Equation 1 is simply based on energy availability and requirements and does not say anything about how many individual processes can be active at any given time – the missing factors of capacity and instantiation or substrate (Section 1.2). An additional factor can be introduced as $f_x$: the fraction of $BOPS_x$ that can be carried out due to the 'supply' of physical instances. In other words, the maximum BOPS will depend on how many molecular processes are 'in the queue' at any time and ready to be driven by available energy at the required level. Formulating this as a fraction of the potential BOPS is one option, with the advantage that it remains bounded by the energy constraints and reduces mismatches that might arise in, for example, comparing the total supply of instances directly against the energy availability and cost. Equation 1 then becomes:

$$BOPS_x \geq f_x \frac{1}{\alpha_x E_{Landauer}} \varepsilon_x P$$

(5)

That can be rearranged to yield:

$$f_x \geq \frac{BOPS_x \alpha_x E_{Landauer}}{\varepsilon_x P}$$

(6)

Assuming that $\alpha_x$ and $\varepsilon_x$ are known (or estimated), as is the environmental $T_0$, for a given $BOPS_x$ and $P$ constraint, then a lower limit is obtained for $f_x$.

Written in this way $f_x$ combines both habitability and occupancy of an environment – two properties that are generally separated in classical approaches. If $f_x = 1$ this implies a process running at maximum capacity across the measured environment (whether a single organism or a planetary biosphere), such that the number of physical instances (opportunities) is precisely matched to the amount and distribution of power to drive those instances to compute. For example, the number of translation amino acid attachments ready to take place is matched to the available chemical energy. Values in the range $\frac{BOPS_x \alpha_x E_{Landauer}}{\varepsilon_x P} \leq f_x < 1$ represent all situations where the available power for a given rate of computation for that process exceeds that required for the number of physical instances. $f_x$ will depend on not just the occupancy of a living system or biome but also time-varying requirements of that system. For example, strong seasonality may impact computation rates via the internal regulation of organisms in response to changing conditions, it could also impact the supply of physical instances. It should be noted that the choice of variables used here is just one option, and there may be advantages to other formulations or in combining variables such as forming an 'instantiation efficiency' given as $\mu_x = f_x \varepsilon_x$.

A quantity of interest is a measure of total habitability (computational potential) and occupancy, which might be represented by a mean value $f_H$:

$$f_H = \frac{1}{N} \sum_{i=1}^{N} f_i$$

(7)

under the assumption of equal weight to all processes in evaluating the success of a living system. This may not be a good, or necessary assumption. As discussed, the BOPS of processes that are vital for a living system's existence are unlikely to all be the same. The BOPS of translation may always be higher than (for example) the BOPS of synaptic transmission across Earth's entire biosphere owing to the proportion of non-synaptic information in single-celled microbial life that might (by sheer number) dominate and boost the planetary BOPS for translation. Nonetheless, the construction of a properly weighted $f_H$ would be a novel, and perhaps a more complete measure of habitability.

## 2.3 Estimates of terrestrial computation

The framework for CZs can be applied at different levels of granularity. An obvious challenge is the need for insight to real BOPS energy requirements and ranges for $f_x$. For a planetary environment the capacity for computation in an extant biosphere may be large. As discussed above, computation may include core operations in DNA and RNA, as well as many modes of information sensing and information communication within and between organisms.

Estimates of computational capacity have been made for specific terrestrial processes, such as nucleotide transcription operations (Landenmark et al., 2015). In that study the number of extant DNA bases on the Earth is estimated as $5.3 \times 10^{37} (\pm 3.6 \times 10^{37})$. With a range of assumptions made on transcription rates (10-40 bases per second) and, assuming that all instances are engaged in processing at a given time ($f_{nucleotide} = 1$), an estimate is arrived at for $BOPS_{nucleotide} < 10^{39} \ s^{-1}$ (Landenmark et al., 2015).

However, if the theoretical minimum energy requirement for each operation is of the order $\sim 10^{-19}$ J at 15° Celsius based on an order-of-magnitude estimate of $\sim 10 \ E_{Landauer}$ (see above, although for nucleotide operations this is likely higher), the upper limit for power utilized by Earth's nucleotide operations is implied to be $10^{20}$ watts, which exceeds the power of absorbed solar radiation at Earth's surface (the dominant power source) by a factor $\sim 10^3$. This also exceeds previous estimates of the biosphere's total power budget (based on REDOX reactions leading to charge separation across semipermeable membranes to drive metabolism, and excluding plant transpiration, Nealson, 1997) of $\sim 100$ terawatts by a factor of $\sim 10^6$.

This clearly implies that the actual $BOPS_{nucleotide}$ for the Earth must be many orders of magnitude less than the estimated upper limit. The most likely reason is that the true fraction of instances engaged in transcription at any moment ($f_{nucleotide}$) is significantly less than $10^{-6}$ (to be consistent with the estimated biosphere power budget). Estimates of single cell median power usage by prokaryotes in low energy environments on the Earth range from $2.23 \times 10^{-18}$ watts per cell for aerobic heterotrophs to $1.50 \times 10^{-20}$ watts per cell for methanogens (Bradley et al., 2020). The average genome size for prokaryotes is $\sim 3.2 \times 10^6$ (Landenmark, 2015) and if all instances were involved in transcription with $\sim 10 \, E_{Landauer}$ energy per operation this would require $\sim 10^{-11}$ watts per cell. In some prokaryote species transcription dominates the energy budget, consuming about 2/3 of cellular ATP (e.g., Russell and Cook, 1995). Therefore, an actual upper limit to the transcription fraction may be $f_{nucleotide} \sim 10^{-9} - 10^{-7}$ to be consistent with the cellular power budget.

Applied globally (and assuming prokaryotes dominate the total DNA count) this suggests that $BOPS_{nucleotide}$ is less than $10^{30} - 10^{32} s^{-1}$, with a power budget of less than ~100 gigawatts to 1 terawatt, consistent with estimates of the total biosphere power.

**2.4 A worked example 1: Generalizing the classical HZ**

Equations 1 and 2 can be examined in the context of phototrophic life around a stellar energy source. Specifically, the scaling of power flow $P$ with environmental (surface) temperature of a planet $T_0$ and thermodynamic efficiency $\varepsilon$ can be cast in terms of planet-star separation $d$. To first order (ignoring the complications of planetary albedo and atmospheric radiative transfer) stellar power per unit area at a planetary surface simply scales as $P = \sigma R_*^2 T^4 / d^2$, where $T$ is the stellar temperature and $R_*$ is the radius of the stellar photosphere. In thermal equilibrium $T_0 = T \left(\frac{R_*}{2d}\right)^{1/2}$ (assuming no planetary reflection, so that the albedo is zero) and Equation 2 provides $\varepsilon = 1 - \frac{4T_0}{3T} + \frac{1}{3}\left(\frac{T_0}{T}\right)^4$. Thus,

the absolute upper limit on photon driven processes (treated as computational in this example) will be:

$$BOPS_{limx} = \frac{(2d)^{1/2}}{\left(a_x k_B T R_*^{1/2} \ln 2\right)} \left[1 - \frac{4}{3}\left(\frac{R_*}{2d}\right)^{1/2} + \frac{1}{12}\frac{R_*^2}{d^2}\right] \frac{R_*^2 \sigma T^4}{d^2}$$

(8)

If we choose $a_x = 10$ for computational processes, the curves for operation rate upper limits are shown in Figure 1 for both a solar mass/radius star of temperature 5800K and a star of mass $0.1 M_\odot$, radius $0.16 R_\odot$, and temperature 2900K commensurate with a zero-age main sequence low-mass M-dwarf (with commensurate stellar luminosities).

The peak computational potential is a factor $8 \times$ higher around the hotter star and occurs at a star-planet separation of only 1.25 solar radii (measured from the stellar center), a distance at which the planet equilibrium temperature would be ~3670 K. This same separation factor applies to the lower mass star, where the peak is again at 1.25 times the stellar radius ($1.25 \times 0.16$ solar radii) for a planet temperature of ~1830K. These are of course extremely high temperatures compared to traditionally considered zones for living systems and simply represent the theoretical maxima in terms of thermodynamic efficiency and peak stellar power per unit area.

Alternatively, and more realistically, the efficiency $\varepsilon$ will be the product of the absolute thermodynamic efficiency with process-specific, environmentally dependent functions, where temperature is likely to be one important factor. Terrestrial biochemistry exhibits a variety of temperature dependencies that can be related to core computational processes. For example, gene expression patterns rates in cells are in general known to change with temperature (e.g., Charlebois et al., 2018). To illustrate this in the present example two models are deployed: a simple, heuristic, temperature dependent efficiency for biological operation rates given by : $r_{gauss}(T) = e^{-\frac{(T_0 - T_{opt})^2}{2 \Delta T_{opt}^2}}$, where $T_{opt}$ is an optimal/peak temperature for biological operations and $\Delta T_{opt}$ is the width of this unit normalized Gaussian efficiency model (yielding $e^{-\frac{\left(T\left(\frac{R_*}{2d}\right)^{1/2}-T_{opt}\right)^2}{2\Delta T_{opt}^2}}$ when substituting for $T_0$ as above), and a terrestrially-inspired model based on the Sharpe-Schoolfield model of biological rates. The Sharpe-Schoolfield model combines the Eyring temperature dependent rate equation for specific biochemical reactions with a parameterized probability for rate-determining enzymes to be active at a given temperature (e.g., see Gibert & De Jong 2001 for the formulation adopted here). This yields a rate $r_{s-s}(T)$ governed by several parameters, the most relevant of which for the present study are the 'reference' temperature at which rate-determining enzymes are maximally active, and the low and high temperatures at which enzymes are 50/50 active/suppressed. We choose these 3 temperatures to be 15, 1, and 40 °C respectively. The expression for $BOPS_{limx}$ is then simply:

$$BOPS_{limx} = \frac{(2d)^{1/2}}{\left(a_x k_B T R_*^{1/2} \ln \ln 2\right)} \left[1 - \frac{4}{3}\left(\frac{R_*}{2d}\right)^{1/2} + \frac{1}{12}\frac{R_*^2}{d^2}\right] r(T) \frac{R_*^2 \sigma T^4}{d^2}$$
(9)

where $r(T)$ is set as either $r_{gauss}(T)$ or $r_{s-s}(T)$. The result of this is straightforward; the $BOPS_{limx}$ is dominated by the form of the process-dependent efficiency factor. Figure 2 plots the computational limits for $r_{gauss}(T)$ with $T_{opt} = 288$K and $\Delta T_{opt} = 15$K, roughly commensurate with optimal conditions for many terrestrial organisms, and the $r_{s-s}(T)$ model described above. The Sharpe-Schoolfield reaction response is asymmetric with respect to temperature (star-planet separation) and skewed towards higher temperatures than the reference temperature due to the Eyring reaction rate temperature dependency. The result for the planet-star separation range with significant computational potential around a 5800 K ad 2900 K star (as above) is therefore (and unsurprisingly) broadly like orbital ranges obtained from classical HZ calculations (e.g., Kasting et al. 1993, Kopparapu et al. 2013). Around the lower mass star ($M = 0.1 M_\odot, R_* = 0.16\ R_\odot$) the photon-driven computation range would be at less than a tenth of the planet-star separation and more than an order of magnitude narrower. It is also apparent in Figure 2 that the thermodynamic efficiency factor causes the peak of computational potential to be ~7% lower around the cooler star compared to the hotter, solar mass star (cf. Scharf, 2018). This of course assumes efficiency differences solely due to the difference in exergy. The actual energy flux at a planetary surface is also dependent on wavelength-dependent atmospheric scattering and absorption. An Earth-similar planet around a low-mass star is expected to have larger irradiance at its surface (for a clear atmosphere) than around a higher mass star (e.g., Kasting et al., 1993).

**2.5 A worked example 2: Dyson structures for computation.**

Another simple example of the application of computational zones is for hypothetical extrapolations of technological energy use. This is particularly relevant to the topic of technosignatures that complement searches for biosignatures and might be more unambiguous, if detected (Wright et al. 2022). One perennially discussed hypothesis is that alien life may have reached an advanced technological stage, where it builds engines enabling it to perform large amounts of computation (Davies, 2012; Shostak, 2018; Becker, 2022). This suggests looking beyond the understood biological qualities of life to detect such situations.

In this kind of scenario total power and efficiency are of paramount importance. In 1960 Dyson introduced an extension to the idea of planetary habitats (Dyson, 1960, also Dyson, 1966) in the limiting case where living systems repurpose matter to build structures capable of (in principle) capturing the entire electromagnetic output of a host star. In effect enclosing the star, most likely with an ensemble of orbital structures, since a monolithic and rigid structure would be inherently unstable (see Wright, 2020). Assuming thermal equilibrium in the habitable range for such structures built around a solar-analog star at a distance of ~1 AU, Dyson suggested the possibility of point-like infrared astronomical sources (with emission wavelengths peaking at $\lambda \lesssim 10 \mu m$) due to the re-emission of stellar energy at higher entropy. In other words: similar to a planet in thermal equilibrium, the total energy absorbed from the star must ultimately be thermalized and re-emitted at the same rate, but at a lower effective temperature (and as a higher entropy radiation field, consisting of more numerous but lower energy photons). As with the previous example for phototrophic life, the absolute thermodynamic limits on the efficiency of Dyson structures (ignoring actual work done) can be examined through the use of photon exergy, and this leads very naturally to a number of insights to the hypothetical configurations of structures supporting computation.

The effective temperature ($T_0$) in thermal equilibrium of a full, spherical structure of radius $R$ (and zero inner albedo) around a star of luminosity $L_* = 4\pi R_*^2 \sigma T_*^4$ (and assuming $T_0 = [L_*/(4\pi\sigma R^2)]^{1/4}$ where $\sigma$ is the Stefan-Boltzmann constant) will be:

$$T_0 = \left(\frac{R_*}{R}\right)^{1/2} T_*$$

(10)

Where it should be noted that the $T_0$ of this structure (assumed to form, in effect, a full spherical enclosure at radius $R$) is systematically higher by a factor $2^{1/2}$ than the equivalent calculation for a planet, since the cross-section of absorption (a spherical surface) is equal to the emission surface area, compared to the situation for a planet that has a geometric cross-section of absorption $\pi R_p^2$ and an emission area $4\pi R_p^2$ - thereby diluting the effective emission temperature for a fixed energy flux.

As discussed in 2.1 an assumption of an isotropic incoming (and outgoing) radiation field is a simplifying approximation, with geometrical corrections that can be applied to the efficiency to account for (e.g.) the finite emission disk of a star (Badescu 2023). Nonetheless, for the sake of illustration here we adopt that approximation; a more thorough study of how work extracted in Dyson structures can be found in Wright 2023. The exergy at the structure's inner surface will be (see Equation 2 above):

$$W_{T,T_0} = \sigma T_*^4 \left(1 - \frac{4}{3}\left(\frac{R_*}{R}\right)^{1/2} + \frac{1}{3}\left(\frac{R_*}{R}\right)^2\right)$$

(11)

So that the maximum possible thermodynamic efficiency $\varepsilon$ for the use of stellar energy in the isotropic approximation is:

$$\varepsilon = 1 - \frac{4}{3}\left(\frac{R_*}{R}\right)^{1/2} + \frac{1}{3}\left(\frac{R_*}{R}\right)^2$$

(12)

i.e., $W_{T,T_0} = \varepsilon \sigma T_*^4$. Equations 10, 11, & 12 hold for all objects whose single-sided emission surface area is equal to their cross-sectional area of absorption (i.e., including partial spherical shells). It is implicitly assumed that emission from the interior does not affect the stellar temperature or the equilibrium temperature, rather that the interior behaves as a cavity blackbody (of zero albedo, see also below) and the only net loss of energy is through the exterior.

Because of the simple $R_*/R$ scaling in Equation 12 it can be immediately seen that at 1 AU around a solar-analog star, for a 'habitable' $T_0 = 290K$, $R \simeq 215R_*$ and $\varepsilon \simeq 91\%$. In Figure 3, $\varepsilon(R)$ is plotted for a range of R/$R_*$. The $R/R_*$ values are also indicated for a Dyson structure of temperature 290 K (i.e., a habitable zone temperature) located around a star of solar temperature (5800 K) and a very cool M-dwarf star (~ 2000 K). With $R/R_*$ of 397 and 48 respectively Figure 3 shows that while a habitable structure around a low-mass star may be significantly smaller (using perhaps a factor $(48/397)^2 \sim 0.015$ less material to construct) it will also be inherently less efficient at 80.7% compared to 93.3% for the hotter star. As previously, these efficiencies are the theoretical maximums, and do not account for other sources of inefficiency, such as in the conversion of photons to electrical or electrochemical work. It is also assumed that all photon energy converted into useful work (driving computation) is eventually thermalized. The 'opacity' of computation would in practice modify the rate of flow of energy and the equilibrium temperature of the structure (akin to a greenhouse effect).

For a solar luminosity $3.826 \times 10^{26}$ W and a 90% efficiency Dyson structure with (for example) $T_0 \sim 290$K, the total possible (upper-limit) rate of irreversible bit-operations occurring at the Landauer limit of energy use is $\sim 10^{47}$ sec$^{-1}$. The number of bit-operations in a digital floating-point operation (FLOP) depends on the computing architecture but could be $\sim 10^3$ (based on 32 bits used for a single-precision floating point number and assuming roughly $32^2$ individual bit operations per FLOP). Consequently, no more than $\sim 10^{44}$ (or 100 tredecillion) FLOPS could be supported by one solar luminosity of power (assuming only thermodynamic efficiency limits at this environmental temperature).

Although analogous to calculating a classical HZ, this example helps illustrate how CZs are a generalization of habitability. First, since computation is, in principle, agnostic to its implementation, it is the efficiency of stellar energy utilization that becomes an important figure of merit. Second, environmental temperature may be a secondary consideration for computation, and since the Landauer limit favors lower operational temperatures, the possibility of hierarchical energy use (or recycling) is indicated.

A Dyson structure around our present-day Sun at 1 AU will be a net absorber of low-entropy solar power and net emitter of high-entropy power: 'waste heat' that cannot directly support computation in biological or digital systems existing at that same blackbody temperature. But there is no reason why higher entropy radiation cannot be further utilized for computation in a colder environment that may be too cold for liquid water and normal terrestrial biological function. The idea of nested stellar energy capture has been previously discussed as "Matrioshka brains" (Bradbury, 1999; see also Sandberg, 1999), with a more thorough treatment given by Wright (2023). The ultimate endpoint of such a system would be a final, outermost structure with an environmental temperature close to that of the cosmic background at ~3K. Consequently, the possibility exists for energy utilization for computation that would result in very low temperature, point-like, far-infrared astrophysical sources (e.g., 50 K implies a Wien's peak of $\sim 50 \mu m$). For comparison Dyson's original concept produces infrared sources at $\lesssim 10 \mu m$. However, the work of Wright (2023) indicates that nested Dyson structures may have limited or no advantage in some circumstances, calling into question the motivations for their construction.

Exactly how the electromagnetic energy available for useful work in computation can be exploited from cooler emission is left as an open question. Recent advances in the conversion of infrared light to electrical charge suggest that there are ways to directly harvest lower-energy photons (see for example work on intermediate wavelengths, Nielsen et al., 2022), and it could be that a heat engine approach would also be workable (Wright 2023), with efficiency losses at least as large as those given by Carnot's theorem: $W = (T_{hot} - T_{cold})/T_{hot}$. In all cases though, a general limit to the *amount* of computation that is possible could be obtained from the Landauer limit.

## 2.6 Sub-stellar computational zones

Substellar objects such as brown dwarfs are seldom considered as viable power sources for habitable environments. With strongly age-dependent $T_{eff} \sim 750 - 2200$ K and luminosities $\sim 10^{-3} - 10^{-5} L_\odot$, the HZ for Earth-analog planets around these objects would be perilously close to the orbital Roche limit (Bolmont et al., 2011) and planets in this orbital range could undergo severe early atmospheric and water loss (Barnes and Heller, 2013). Other potential challenges for living systems in planetary environments around brown dwarfs include a lack of UV radiation that could drive some prebiotic chemistry and a lack of photosynthetically active (shorter wavelength) radiation (e.g., Lingam et al. 2020).

But as electromagnetic power sources for driving differently instantiated computation, substellar objects offer some intriguing possibilities. With a brown dwarf radius of $\sim 1 R_{Jupiter}$, a 90% efficiency Dyson structure could be built with a radius close to 0.1 AU - requiring only 1% of the material of a 1 AU structure and resulting in an emission temperature of $\sim 53 - 156$ K depending on the brown dwarf temperature (see above). With total power $\sim 10^{21} - 10^{23}$ W, and assuming bit-operations at the Landauer limit, a cool brown dwarf could nonetheless support a total computational capacity of at least $10^{19}$ times Earth's present net technological FLOPS level.

Substellar objects surrounded by energy capture structures would represent a previously unconsidered population of point-source, far infrared emitters (~$50 - 100 \mu$m) with no optical counterparts - much like Dyson's original 1960 proposal but shifted to longer wavelengths. Existing searches (e.g., Carrigan 2009) have generally focused on shorter infrared wavelengths. While largely complete, it is worth noting that surveys such as the IRAS Point Source Catalog and Faint Source Catalog at 12, 25, 60, and 100 $\mu$m still include at least several dozen unidentified sources (e.g., Rowan-Robinson, 2021) and with ~1 arcmin resolution at 60 $\mu$m determining counterparts can be challenging. The Spitzer Archival Far-Infrared Extragalactic Survey at 70 and 160 microns contains some 27,500 and 14,600 sources respectively in a total of 180 square degrees that are presumed extragalactic in origin (Hanish et al., 2015), but could contain galactic sources as described here. The AKARI (ASTRO-F) space telescope produced a catalog (with the Far-Infrared Surveyor, FIS instrument) of ~430,000 sources using fluxes from six bands between 9 $\mu m$ and 180 $\mu m$ (Yamamura et al., 2010) and could be searched for objects without counterparts at other wavelengths.

## 3. Conclusion

Shifting focus towards the piecewise *processes* of matter involved with life, articulated as computation, offers a natural way to move beyond the traditional concept of an astrophysical (or geophysical) habitable zone, towards a more universal and predictive framework. Work that has been carried out on habitability through characteristics such as REDOX potentials in an environment, or nutrient and elemental availability limits already takes a step in this direction (e.g., Hoehler, 2007; Hoehler et al. 2007; Higgins and

Cockell, 2020). But computation – as defined here in the systematic manipulation of information encoded in physical states following a set of rules – may provide a more generally applicable way to combine estimates of the possibility of life and the occupancy of life, with fundamental energetic constraints from Shannon entropy and the von Neumann-Landauer limit.

In this present work the focus is on three, simplified limiting factors for computation: capacity, energy, and instantiation or substrate. For comparison, the classical orbital HZ is typically derived from energy flow and planetary properties linked to only some of life's requirements. A first-principles upper limit on biological operations per unit time applied to the Earth suggests a global rate of DNA transcription nucleotide operations of $< 10^{30} - 10^{32} s^{-1}$ to be consistent with the biosphere's estimated power budget and that of individual cells.

A parameterization of biological computational dependencies on factors such as temperature is indicated as necessary for calculating computational zones. This is a departure from the usual focus in circumstellar HZs on liquid water, or on metabolic rates, but for terrestrial biology at least, computational environmental dependency is likely to track closely those same underlying metrics. Future work should be done to examine the structure of the hierarchy of computational and 'non-computational' processes that can be equated with living systems, and the partitioning of energy flow and matter within this hierarchy. It is also suggested that laboratory work on quantifying the functional form and calibration of these environmental dependencies (e.g., operation rates versus temperature) across specific biological computational processes would be of interest.

Examining the hypothetical example of technological stellar energy capture structures through the lens of computation also provides insight to potential strategies for optimizing thermodynamic efficiency and total power capture. Nested capture structures are an interesting possibility in some circumstances (Wright 2023), as is the utilization of sub-stellar objects as power sources. In both scenarios it is predicted that a population of point-like far-infrared emitters ($\sim 50 - 100 \mu$m) with no optical counterparts could be an observable consequence.

A highly speculative (not entirely original) corollary to the work presented here is that technological species might eventually learn that the joint capacity, energy, and instantiation requirements of biological computation overly restrict the availability of suitable environments (or lead to asymptotic burnout, Wong and Bartlett, 2022). Physical laws, as shown by the thermodynamic limits on efficiency, may incentivize transitioning to other computational modes of existence.

Computation is robust yet constrained in our universe. Understanding and quantifying those constraints through the computational zones approach proposed in this paper may provide new clarity in the search for living systems, even in the event of them taking very different forms.

**Credit Statements**

Caleb Scharf is responsible for the initial conception and major structural elements and primary writing of this article. Olaf Witkowski is responsible for additional conceptual elements and writing forming a significant part of this article.

**Acknowledgements**

The authors thanks Piet Hut, Michael Wong, and Chris Kempes for helpful discussions, and hope that they get away with publishing the penultimate paragraph.

**Authors' Disclosure Statement**

Authors declare no conflict of interest.

**Funding Statement**

This work was supported in part by the NASA Astrobiology Program through participation in the Nexus for Exoplanet System Science and NASA Grant NNX15AK95G.

Zhu X-G, Long SP, and Ort DR. What is the maximum efficiency with which photosynthesis can convert solar energy into biomass?. *Current Opinion in Biotechnology*. 2008; 19, 153

**FIGURES**

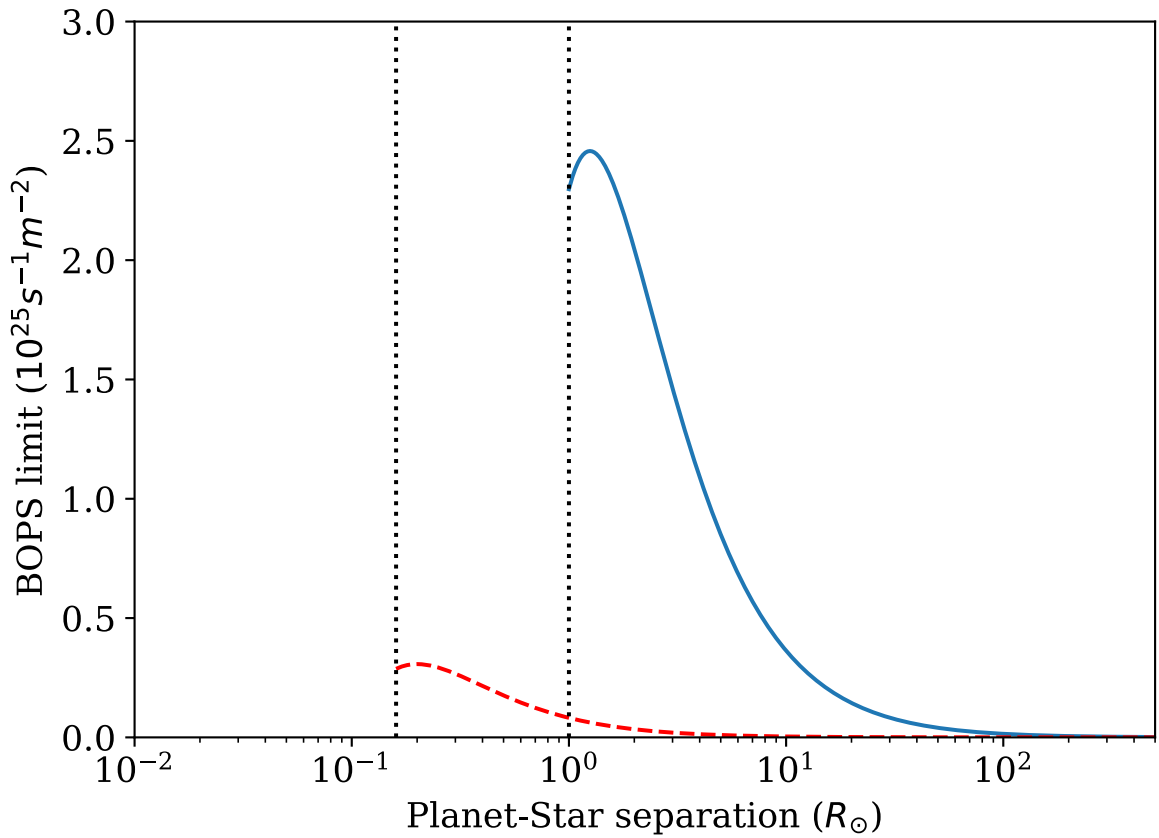

**Figure 1:** *Upper limits on photon driven biological operations per second per square meter on a planetary surface are plotted as a function of planet-star separation (in units of solar radius) according to Equation 8 for two stellar types: a solar analog (Right solid curve: $1M_\odot, 1R_\odot, T = 5800K$) and a low-mass M-dwarf (Left dashed curve: $0.1M_\odot, 0.16R_\odot, T = 2900K$). Vertical dotted lines indicate the stellar radius in either case.*

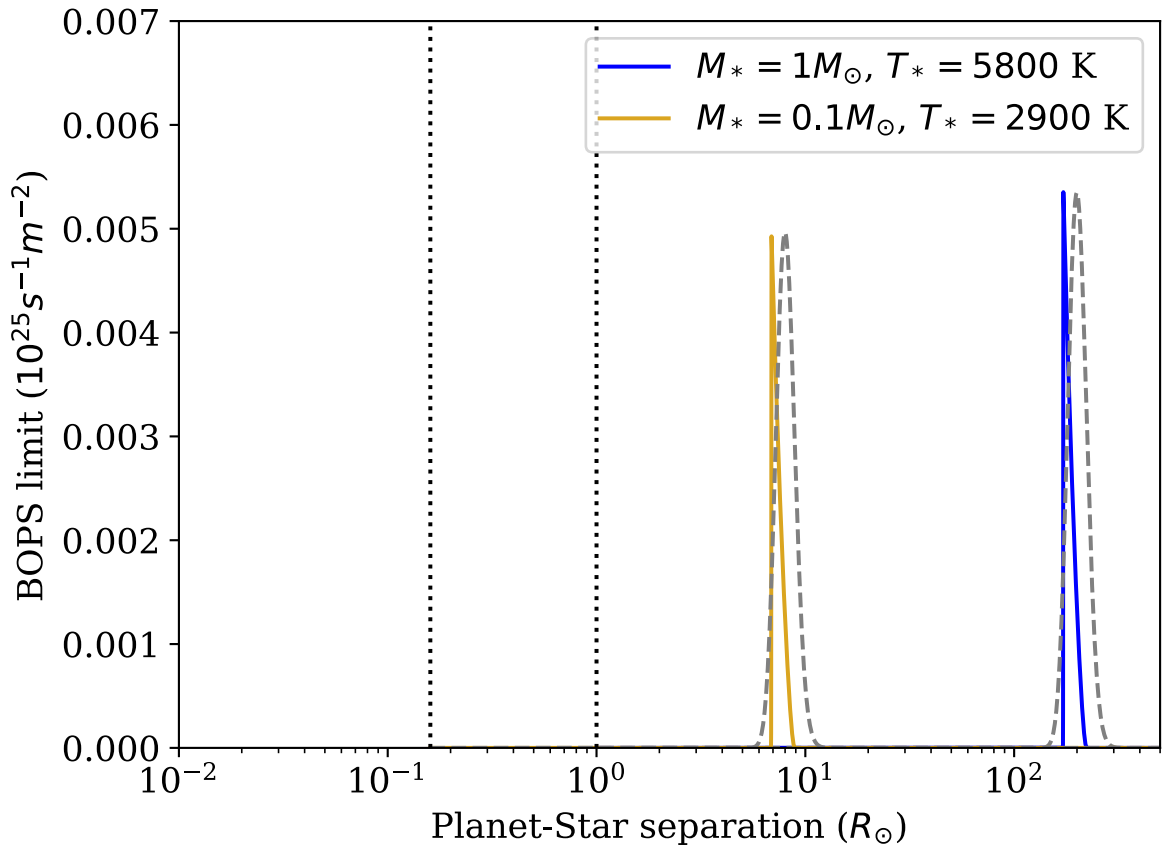

**Figure 2:** *Upper limits on photon driven biological operations per second per square meter on a planetary surface are plotted as a function of planet-star separation (in units of solar radius) according to Equation 9 that incorporates two models for the temperature dependence of the efficiency of biological operations in addition to thermodynamic efficiency constraints. Dashed curves correspond to a simple heuristic (gaussian) model, solid curves correspond to a Sharpe-Schoolfield model, both with a reference or optimal temperature set at 15 Celsius. Results are plotted for a solar analog (Right solid and dashed curves: $1M_\odot, 1R_\odot, T = 5800K$) star and a low-mass M-dwarf (Left solid and dashed curves: $0.1M_\odot, 0.16R_\odot, T = 2900K$). Vertical dotted lines indicate the stellar radius in either case.*

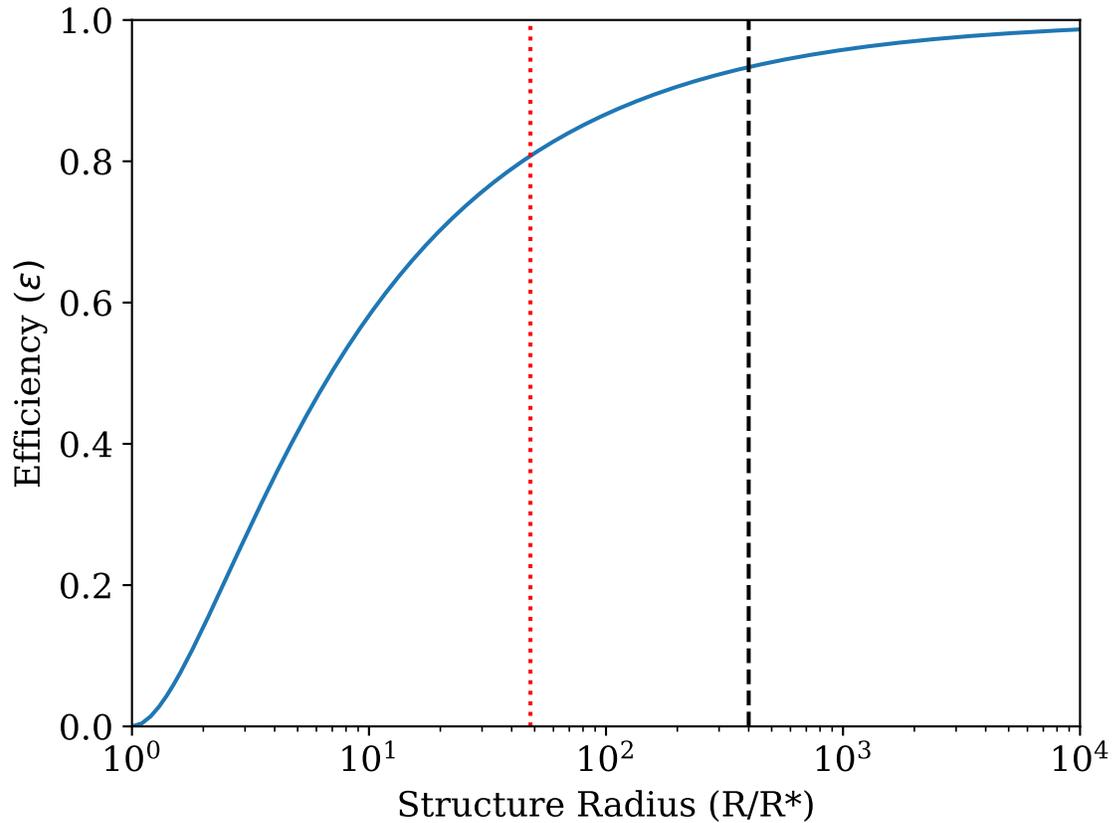

**Figure 3:** *The maximum possible thermodynamic efficiency (ε, using the isotropic approximation) of the utilization of stellar energy by spherical Dyson structures (full or partial) as a function of structure size (radius) scaled to the stellar radius. Vertical lines represent (left, dotted) the required size of a 290K 'habitable' Dyson structure around a low mass, cool M-dwarf star of $T_* = 2{,}000K$ and (right, dashed) around a solar-mass star with $T_* = 5{,}778K$*